\documentclass[12pt,a4paper]{article}
\usepackage{a4wide}
\usepackage{epsfig}
\usepackage{cite}
\usepackage{amsmath}
\usepackage{graphicx}
\usepackage{latexsym}
\usepackage{array}

\newlength{\absize}
\newcommand{\Lumint}{{\cal L}_{\rm int}}

\newcommand{\dd}{\mbox{{\rm d}}}

\def\lsim{\mathrel{\rlap{\raise 2.5pt \hbox{$<$}}\lower 2.5pt
\hbox{$\sim$}}}

\begin{document}

\thispagestyle{empty}
\renewcommand{\thefootnote}{\fnsymbol{footnote}}
\newpage\normalsize
\pagestyle{plain} \setlength{\baselineskip}{4ex}\par
\setcounter{footnote}{0}
\renewcommand{\thefootnote}{\arabic{footnote}}
\newcommand{\preprint}[1]{%
\begin{flushright}
\setlength{\baselineskip}{3ex} #1
\end{flushright}}
\renewcommand{\title}[1]{%
\begin{center}
\LARGE #1
\end{center}\par}
\renewcommand{\author}[1]{%
\vspace{2ex} {\Large
\begin{center}
\setlength{\baselineskip}{3ex} #1 \par
\end{center}}}
\renewcommand{\thanks}[1]{\footnote{#1}}
\renewcommand{\abstract}[1]{%
\vspace{2ex} \normalsize
\begin{center}
\centerline{\bf Abstract}\par \vspace{2ex}
\parbox{\absize}{#1\setlength{\baselineskip}{2.5ex}\par}
\end{center}}

\vspace*{4mm} 

\title{The Higgs-like boson spin from the center-edge asymmetry in the diphoton
channel at the LHC} \vfill

\author{P. Osland,$^{a,}$\footnote{E-mail: per.osland@ift.uib.no}
A. A. Pankov$^{b,}$\footnote{E-mail: pankov@ictp.it}
  and
A. V. Tsytrinov$^{b,}$\footnote{E-mail: tsytrin@rambler.ru}}
\begin{center}
$^{a}$Department of Physics and Technology, University of Bergen, Postboks 7803, \\
N-5020  Bergen, Norway\\
$^{b}$The Abdus Salam ICTP Affiliated Centre, Technical
University of Gomel, \\ 246746 Gomel, Belarus
\end{center}
%
%
%
\vfill \abstract{We discuss the discrimination of the 125 GeV
spin-parity $0^+$ Higgs-like boson  observed at the LHC, decaying into two photons,
$H\to\gamma\gamma$, against the hypothesis of a minimally coupled
$J^{P}=2^+$ narrow diphoton resonance with the same mass and giving
the same total number of signal events under the peak.
We apply, as the basic observable of the analysis, the
center-edge asymmetry $A_{\rm CE}$ of the cosine of the polar
angle of the produced photons in the diphoton rest frame to
distinguish between the tested spin hypotheses. We show that
the center-edge asymmetry $A_{\rm CE}$  should provide strong
discrimination between the possibilities of spin-0 and spin-2 with
graviton-like couplings, depending on the fraction of $q\bar{q}$ production of the
spin-2 signal, reaching a $\text{CL}_s<10^{-6}$ for $f_{qq}=0$.
Indeed, the $A_{\rm CE}$ has the potential to do better than existing analyses for $f_{qq}<0.4$.
}
\vspace*{20mm} \setcounter{footnote}{0} \vfill

\newpage
\setcounter{footnote}{0}
\renewcommand{\thefootnote}{\arabic{footnote}}

\section{Introduction} \label{sect:introduction}

In 2012 the ATLAS and CMS collaborations announced the discovery
of a new 125~GeV
resonance~\cite{Aad:2012tfa,Chatrchyan:2012ufa} in their search
for the Standard Model (SM) Higgs boson ($H$). It was a great
triumph of the LHC experiments as in all its properties it appears
just as the Higgs boson of the SM. Signals have been
identified in various channels, in particular $H\rightarrow \gamma\gamma$,
$H\rightarrow ZZ^{*}$ and $H\rightarrow WW^{*}$. The next step is
to have precision measurements as well as determinations of the
particle properties, such as its spin, $CP$, decay branching
ratios, couplings with SM particles, and self-couplings.

The inclusive two-photon production process at the LHC,
\begin{equation}
p+p\to \gamma \gamma +X,
\label{proc}
\end{equation}
is considered a powerful testing ground for the SM, in
particular as a discovery channel for Higgs boson
searches. Since the observation of the Higgs-like peak  by both
the ATLAS and CMS experiments, much effort has been
devoted to the comparison (with increased statistics) of the
properties of this particle with the SM predictions for the Higgs
boson, in particular to test the spin-0 character, see
Refs.~\cite{Aad:2013xqa,ATLAS:2013xla,Chatrchyan:2013mxa,Khachatryan:2014ira,Aad:2014eha,Khachatryan:2014kca} 
where the data sets at $\sqrt s=7$ and 8~TeV have been
employed. In this regard, the decay channel in (\ref{proc}) is
particularly suited, because the exchange of spin-1 is excluded, as
the Landau-Yang theorem \cite{Landau:1948kw,Yang:1950rg} forbids a direct decay of an
on-shell spin-1 particle into $\gamma\gamma$, and only spin-2
remains as a competitor hypothesis.

Recent measurements~\cite{ATLAS:2013xla,ATLAS:2013mla,Aad:2013xqa,Chatrchyan:2013mxa,Khachatryan:2014ira,Khachatryan:2014kca}
favor spin-0 over specific spin-2 scenarios. In
particular,  measurements of the spin of the resonance exclude a
minimal coupling of the spin-2 resonance produced through gluon
fusion in the $\gamma\gamma$ channel at almost $3\sigma$, and
approximately at $2\sigma$ in the $ZZ$  and $WW$
channels~\cite{Aad:2013xqa}.

Many proposals have been put forward to discriminate between the
spin-0 and spin-2 hypotheses basically focusing on kinematic
distributions, e.g, angular
distributions~\cite{Choi:2002jk,Gao:2010qx,DeRujula:2010ys,Englert:2010ud,Bolognesi:2012mm,
Choi:2012yg,Englert:2012xt,Banerjee:2012ez,Modak:2013sb,Boer:2013fca,Frank:2013gca},
event shapes~\cite{Englert:2013opa} as well as other
observables~\cite{Boughezal:2012tz,Ellis:2012xd,Alves:2012fb,Geng:2012hy,Ellis:2012jv,Djouadi:2013yb}.
Among the latter, an interesting possibility to discriminate
between the spin hypotheses of the Higgs-like particle was
studied in \cite{Alves:2012fb,Ellis:2012jv} by means of the
center-edge asymmetry $A_{\rm CE}$ where its high potential as a
spin discriminator was demonstrated.

The center-edge asymmetry was first proposed in
\cite{Osland:2003fn,Dvergsnes:2004tc,Dvergsnes:2004tw} for spin
identification of Kaluza-Klein gravitons at the LHC. The
approach based on $A_{\rm CE}$ was further developed in subsequent
papers \cite{Osland:2010yg,Osland:2009tn,Osland:2008sy,Kumar:2011yta} for
 spin identification of heavy resonances in dilepton and diphoton
 channels at the LHC.

Here, we review the application of $A_{CE}$ to the angular study
 of the diphoton production process (\ref{proc}) at ATLAS
extending the analysis done in \cite{Alves:2012fb,Ellis:2012jv} by
accounting for various admixtures of the $gg$ and $q\bar{q}$ production
modes. Also, an optimization of the center-edge asymmetry on the
kinematical parameter which divides the whole range of
$\cos\hat\theta$ into center and edge regions will be
performed in order to enhance the potential of $A_{CE}$ as a
discriminator of spin hypotheses of Higgs-like resonances.

\section{Center-edge asymmetry } \label{sect:ace}

The spin-2 resonance can be produced either via gluon fusion
($gg$) or via $P$-wave quark-antiquark annihilation ($q\bar{q}$).
As we will show, the discrimination between the spin hypotheses is
weakened if the spin-2 particle is produced predominantly via
quark-antiquark annihilation.

In the diphoton decay of a Higgs-like boson, $H\to\gamma\gamma$, the
spin information is extracted from the distribution in the polar
angle $\hat\theta$ of the photons with respect to the $z$-axis of
the Collins-Soper frame \cite{Collins:1977iv}.  A scalar, spin-0, particle
decays isotropically in its rest frame; before any acceptance
cuts, the angular distribution $\dd N^{\rm spin-0}/\dd z$
($z\equiv\cos\hat\theta$) is flat and the normalized distribution can
be written as\footnote{A related issue is the separation of a scalar and a pseudoscalar. Since the two-body pseudoscalar decay distribution would also be flat, the present asymmetry is not useful for this distinction. However, a suitable four-body final state can provide this distinction \cite{Nelson:1986ki,Skjold:1993jd,Godbole:2007cn}.}
\begin{equation}\label{Eq:dS}
\frac{1}{N^{\rm spin-0}}\, \frac{\dd N^{\rm spin-0}}{\dd z} =
 \frac{1}{2}.
\end{equation}
The correspondence between spin and angular distribution is quite
sharp: a spin-0 resonance determines a flat angular distribution,
whereas spin-2 yields a quartic distribution that can be conveniently
written in a self-explanatory way as \cite{Cheung:1999ja,Eboli:1999aq}
\begin{equation}\label{Eq:dgg}
\frac{1}{N^{\rm spin-2}_{gg}}\, \frac{\dd N^{\rm spin-2}_{gg}}{\dd
z} = \frac{5}{32}\,(1+6\,z^2+z^4)
\end{equation}
for the gluon fusion production mode of a spin-2 particle in a
Kaluza-Klein model with minimal couplings and
\begin{equation}\label{Eq:dqq}
\frac{1}{N^{\rm spin-2}_{qq}}\, \frac{\dd N^{\rm spin-2}_{qq}}{\dd
z} = \frac{5}{8}\,(1-z^4)
\end{equation}
for the quark-antiquark annihilation. From Eqs.~(\ref{Eq:dgg}) and
(\ref{Eq:dqq}), the normalized differential distribution for
a spin-2 tensor particle reads
\begin{equation}\label{Eq:dG}
\frac{1}{N^{\rm spin-2}}\, \frac{\dd N^{\rm spin-2}}{\dd z} =
\frac{5}{32}\,(1+6\,z^2+z^4)(1-f_{qq})+
\frac{5}{8}\,(1-z^4)\,f_{qq},
\end{equation}
where $N^{\rm spin-2}=N^{\rm spin-2}_{gg}+N^{\rm spin-2}_{qq}$ and
we denote $f_{qq}=N^{\rm spin-2}_{qq}/{N^{\rm spin-2}}$.
Note that $f_{qq}$ refers to an event fraction, directly proportional to a ratio involving convolution integrals.

The background, which is dominated by the irreducible non-resonant diphoton production,
turns out to be rather large before selection cuts. It is
peaked in the forward and backward directions due to $t$- and
$u$-channel exchange amplitudes. Determining this distribution
precisely from the data is a key challenge of the analysis.
Several methods have been proposed to solve that problem
\cite{Alves:2012fb,Ellis:2012jv,ATLAS:2013xla}.

In practice the shapes in Eqs.~(\ref{Eq:dS}) and
(\ref{Eq:dG}) will be significantly distorted by experimental selection
cuts, resolutions and contamination effects from
background subtractions. However, detector cuts are not taken into
account in the above Eqs.~(\ref{Eq:dS}) and (\ref{Eq:dG}). We will
use these expressions for illustration purposes, in order to
better expose the most important features of the method we use.
The final numerical results, as well as the relevant figures that
will be presented in what follows refer to the full calculation,
with detector cuts taken into account.

We introduce the center-edge asymmetry to quantify the separation
significance between spin-0 and spin-2 resonances following the
definition given in
Refs.~\cite{Osland:2003fn,Dvergsnes:2004tc,Dvergsnes:2004tw,Osland:2008sy,Osland:2009tn,Osland:2010yg,Kumar:2011yta}
for the case of dilepton  and diphoton hadronic production:
\begin{equation}
\label{Eq:ace}
A_{\rm{CE}}=\frac{N_C-N_E}{N_C+N_E}=\frac{N_C-N_E}{N},
\end{equation}
where $N_C$  is the number of events lying within the center range
$-z^*\le z \le z^*$ and $N_E$ the number of events outside
this range (in the edge range). Here, $0<z^*<1$ is a
threshold that can be optimised a priori for
the best separation between spin hypotheses. For instance, in
Refs.~\cite{Alves:2012fb,Ellis:2012xd} it is taken to be
$z^*=0.5$. The interest of this observable should be that,
being defined as a ratio between cross sections, theoretical
uncertainties related to the choice of parton distributions and
factorization/renormalization point should be minimized, and the
same could be true, for example, of the systematic uncertanties on
signal and background normalizations \cite{Alves:2012fb}.

The formulae for $A_{\rm{CE}}$ can be easily obtained from its
definition (\ref{Eq:ace}) and the expressions for the angular
distributions (\ref{Eq:dS}) and (\ref{Eq:dG}):
\begin{equation}\label{Eq:ACES}
A_{\rm CE}^{\rm spin-0}=2\,z^*-1,
\end{equation}
and for the spin-2 case one reads
\begin{equation}\label{Eq:ACEG}
A_{\rm CE}^{\rm spin-2}=f_{qq}\,A_{{\rm CE},qq}^{\rm spin-2}+
(1-f_{qq})\,A_{{\rm CE},gg}^{\rm spin-2},
\end{equation}
where
\begin{equation}\label{Eq:ACEqq}
A_{{\rm CE},qq}^{\rm spin-2}=\frac{1}{2}\,z^*\,(5-{z^*}^4)-1,
\end{equation}
\begin{equation}\label{Eq:ACEgg}
A_{{\rm CE},gg}^{\rm
spin-2}=\frac{5}{8}\,(z^*+2{z^*}^3+\frac{{z^*}^5}{5})-1.
\end{equation}

To evaluate $A_{\rm CE}$ one needs the angular distributions of
the diphoton events relevant to the particular experiment at the
LHC.  Such normalized $\cos\hat\theta$ distributions (simulations)
were presented by ATLAS (Fig.~5 in Ref.~\cite{ATLAS:2013xla}), after
background subtractions and including cuts, hadronization and
detector effects (which are different for the spin-0 and the
spin-2 signal), together
with the observed distribution from background events in the
invariant-mass sidebands ($105$ GeV $< m_{\gamma\gamma} < 122$ GeV
and $130$ GeV $< m_{\gamma\gamma} < 160$ GeV) \cite{ATLAS:2013xla}.
Also, Fig.~2 of Ref.~\cite{Aad:2013xqa} shows the expected
(absolute) distributions of background-subtracted data in the
signal region as a function of $\cos\hat\theta$ for spin-0 and
spin-2 signals. It turns out that for the ATLAS experiment,
using $20.7~\text{fb}^{-1}$ at $\sqrt{s}=8~\text{TeV}$, the
number of background events is about 14300  \cite{Aad:2013xqa}
and the fitted
Higgs boson signal that corresponds to the selection cuts of the  ATLAS
analysis and identification efficiency of photons is about 670 events.\footnote{The expected (NNLO-NNLL QCD) SM number of events is lower, about 390 events. We comment on this discrepancy in the context of Fig.~\ref{fig5} below.}
From a comparison of these
distributions with those described by Eqs.~(\ref{Eq:ACES}) and
(\ref{Eq:ACEG}) for the idealized case one can appreciate the role
of imposing the experimental selection cuts, resolutions and
contamination effects from background subtractions on the
distortion of the idealized pattern and conclude that it is
substantial.

\begin{figure}[htb] 
\vspace*{0.1cm} \centerline{ \hspace*{-2.0cm}
\includegraphics[width=12.0cm,angle=0]{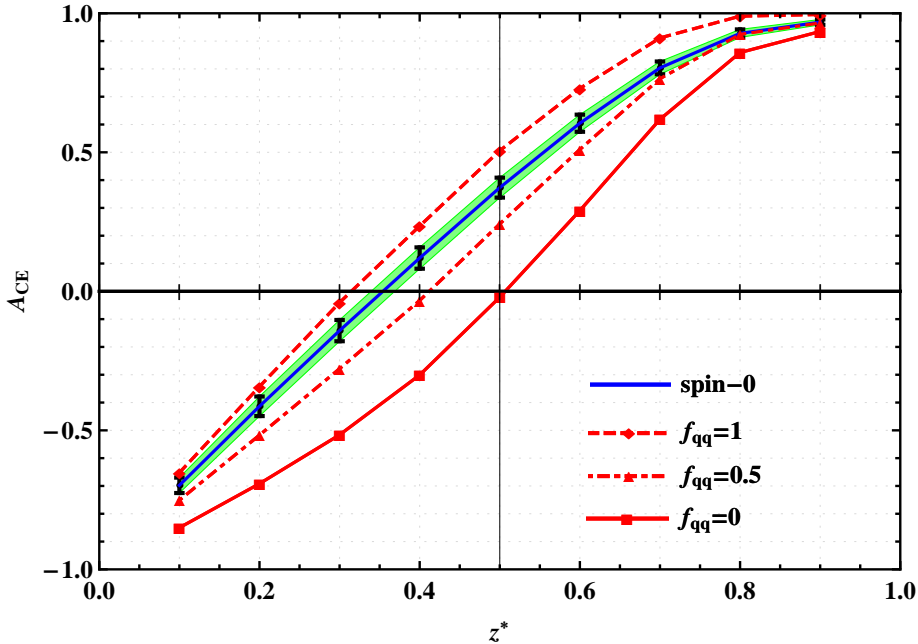}}
\caption{\label{fig1} $A_{CE}$ as a function of $z^*$ for spin-0
(solid, blue, with error bars) and spin-2 hypotheses at different
$f_{qq}$ (red, marked by squares and triangles) from the process
(\ref{proc}) at ATLAS with $\sqrt{s}=$8 TeV,
$\Lumint=20.7~\text{fb}^{-1}$. The vertical bars and corresponding
green band represent 1$\sigma$ statistical uncertainties on
$A_{CE}$ \cite{Alves:2012fb,ATLAS:2013xla},
based on the observed number of events discussed in the text.}
\end{figure}

In Fig.~\ref{fig1} we show $A_{CE}$ as a function of $z^*$ for
spin-0 and spin-2 for different fractions of the sub-processes
($f_{qq}$ = 0, 0.5 and 1) obtained from the distributions depicted
in Fig.~5 of Ref.~\cite{ATLAS:2013xla}. 
The center-edge asymmetry and the corresponding statistical uncertainties
attached to the line of the spin-0 case shown in Fig.~\ref{fig1} were obtained
following the  ``sWeight'' technique developed in \cite{Alves:2012fb} that allows to perform
excellent signal versus background separation.  One should notice that
$A_{CE}$ and its statistical uncertainty depicted in Fig.~\ref{fig1} for the spin-0 and
spin-2 cases  at  $z^*=0.5$  (and with $f_{qq}=0$ for spin 2) are quite
consistent with those derived in Ref.~\cite{Alves:2012fb} (see Table 1).
The figure indicates that
the maximal differentiation of the observable between the
different spin hypotheses occurs at $z^*\approx 0.4-0.6$. It
should be noted that for $f_{qq}$ around 0.75, the $A_\text{CE}$
observable  becomes useless. Typical models, however, like the
Randall--Sundrum model \cite{Randall:1999ee}, favor much lower
values of $f_{qq}$, where the discriminination is substantial.

One should note that systematic uncertainties affecting the signal
yield as a multiplicative factor cancel in the asymmetry $A_{CE}$.
This holds for systematics on luminosity, $z$-independent
selection efficiencies, theoretical errors from renormalization
and factorization scale uncertainties etc. But some types of
errors on an asymmetry measure (e.g. parton distribution function
uncertainties, PDFs) do not cancel. A systematic error of about
3\% comes from PDFs, which does not cancel in the $A_{CE}$
\cite{Alves:2012fb}, was taken into account in the numerical
analysis.

The $A_{CE}$ asymmetry obeys a Gaussian distribution with mean
$\bar A_{CE}$ and standard deviation $\bar\sigma_{A_{CE}}$, which
can be written as
\begin{equation}
\bar\sigma_{A_{CE}} = \sqrt{(1-{\bar A_{CE}}^2)/N}.
\end{equation}
To evaluate the confidence level at which the
spin-2 hypothesis can be excluded we start from the assumption
that spin-0 favors the experimental data as there is strong
motivation for prioritizing the spin-0 hypothesis. In Fig.~\ref{fig1},
the vertical bars attached to the solid (spin-0) line
represent, again as an example, the 1$\sigma$ statistical
uncertainty on $A_{CE}$ corresponding to the Higgs boson
signal events. Comparison of the $A_{CE}$ difference between the spin-0 and
spin-2 curves with the statistical uncertainties allows to make a simple
approximate evaluation of the separation significance of the two spin
hypotheses. 
In this analysis we adopt the assumption that the spin-2 resonance has
the same mass and width as the Higgs boson, and the cross section for
the production and decay of a tensor resonance is normalized by the SM
Higgs rates.
For example, for $f_{qq}=0$ and $z^*=0.5$ one obtains
a separation significance of 6 to $8\sigma$.

\begin{figure}[htb] 
\vspace*{0.1cm} \centerline{ \hspace*{-2.0cm}
\includegraphics[width=12.0cm,angle=0]{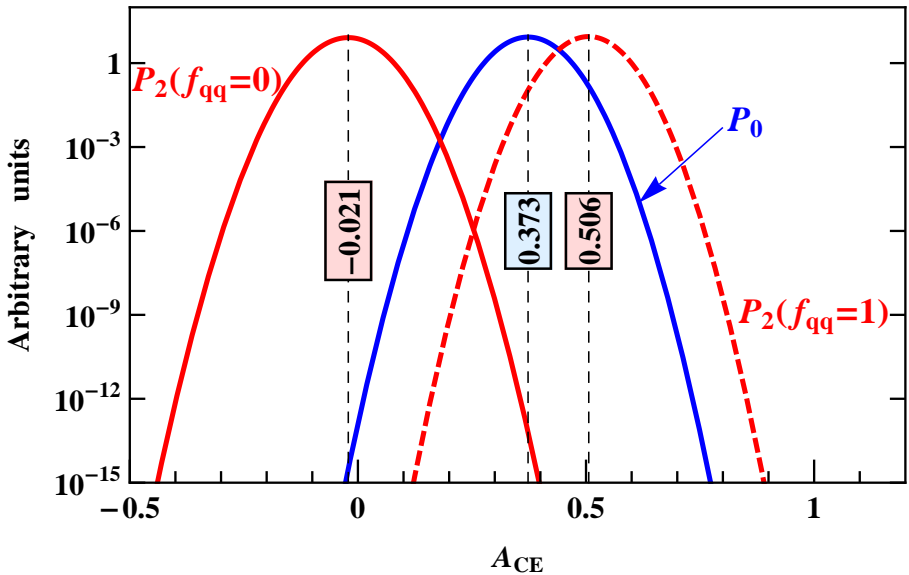}}
\caption{\label{fig2} Probability density functions of the signal center-edge
asymmetry $A_{CE}$ for spin-0 and spin-2 distributions. The $A_{CE}$ values at which the probabilities are maximized, are given in the boxes.}
\end{figure}

In Fig.~\ref{fig2} we show the probability density functions (pdf)
for the hypotheses considered above.
From the integration of the probability density functions shown in
that figure one can calculate $p$-values for rejection of a
hypothesis with tensor resonance.  Then, one should convert the
obtained  $p$-value to the number of standard deviations
($\sigma$'s) as \cite{Alves:2012fb,Beringer:1900zz}
\begin{equation}
Z(\sigma) = \Phi^{-1}(1-p) = \frac{\vert{\bar A_0}-{\bar A_2}\vert
}{{\bar \sigma_0}}, \label{signif}
\end{equation}
where we denote $A_0= A_{CE}^{\rm spin-0}$ and $A_2= A_{CE}^{\rm
spin-2}$, the inverse of the cumulative distribution function of
the standard normal, $\Phi^{-1}(1-p)$, calculated at $1-p$, gives
the standard confidence level  $Z(\sigma)$ of the test in units of
the standard deviation of the Gaussian distribution
\cite{Beringer:1900zz}. Furthermore, $\bar \sigma_0$ here refers to $\bar\sigma_{A_{CE}}$ defined above, evaluated for the spin-zero case.

\begin{figure}[htb] 
\vspace*{0.1cm} \centerline{ \hspace*{-2.0cm}
\includegraphics[width=12.0cm,angle=0]{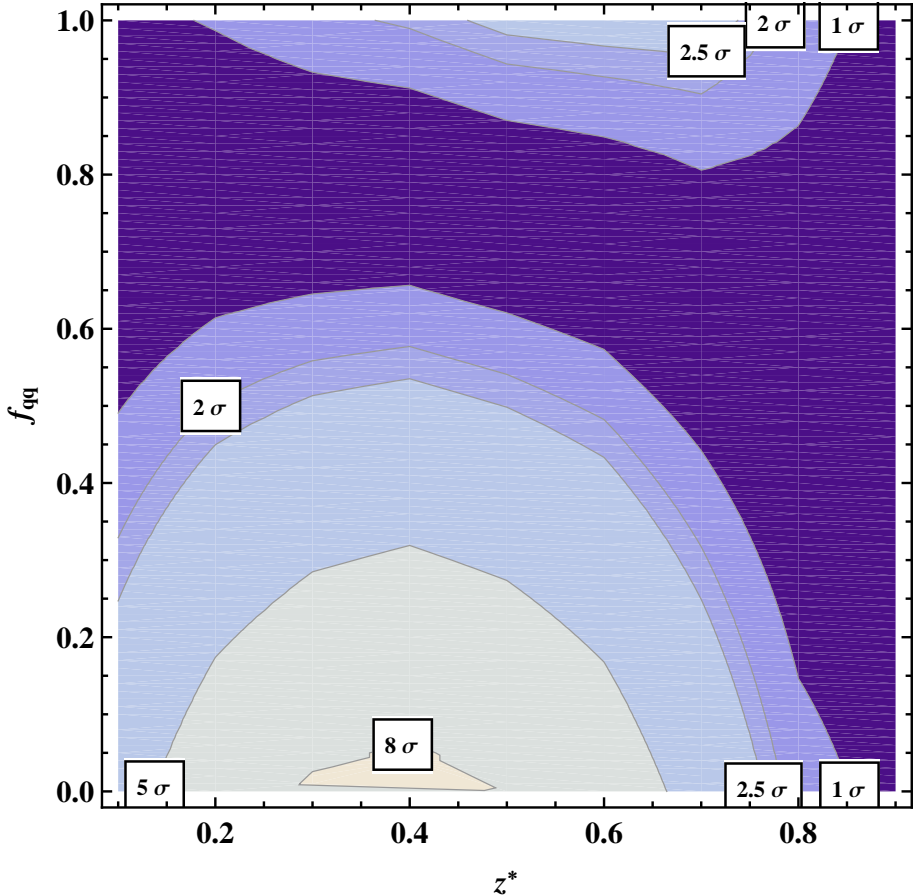}}
\caption{\label{fig3} The significance $Z(\sigma)$ for
spin-0 vs spin-2 hypotheses in the plane of ($z^*$,\,$f_{qq}$),
based on $A_{CE}$ determined from ATLAS simulations for the process (\ref{proc}) at the LHC
with $\sqrt{s}=$8 TeV and $\Lumint=20.7~\text{fb}^{-1}$ \cite{ATLAS:2013xla}. }
\end{figure}

The center-edge  asymmetry here depends on two parameters, namely
the kinematical parameter $z^*$ and the fraction $f_{qq}$ of the
$q\bar{q}$ production of the spin-2 particle. In Fig.~\ref{fig3}
we show a contour plot in the ($z^*$,\,$f_{qq}$) plane for
the separation significance $Z(\sigma)$ defined in
Eq.~(\ref{signif}) and translated into $n$ standard deviations
 attached to the curves, for spin-0 vs spin-2 hypotheses.
Fig.~\ref{fig3} shows that one can optimize the kinematical
parameter $z^*$ in order to obtain the largest separation
significance. In fact, the most suitable $z^*$ is in the
range $z^*=0.4-0.6$. Such optimization can be
applied for the spin separation analysis within the whole
range of values for the fraction $f_{qq}$.

Also, Fig.~\ref{fig3} shows the area
(dark blue) with the smallest separation between the spin-0 and spin-2
signals which occurs for example, for $f_{qq}\approx 0.7$.
In this area of  smaller separation power, the method does not
allow the exclusion of the spin-2 hypothesis when the Higgs-like
boson is produced partially by $q\bar{q}$ annihilation. The
reason is that with this admixture, the sum of the spin-2 $A_{CE}$
pdf's associated to gluon fusion  and quark-antiquark production
is very similar to that of spin-0.
\begin{figure}[tbh!] 
\vspace*{0.1cm} \centerline{ \hspace*{-2.0cm}
\includegraphics[width=12.0cm,angle=0]{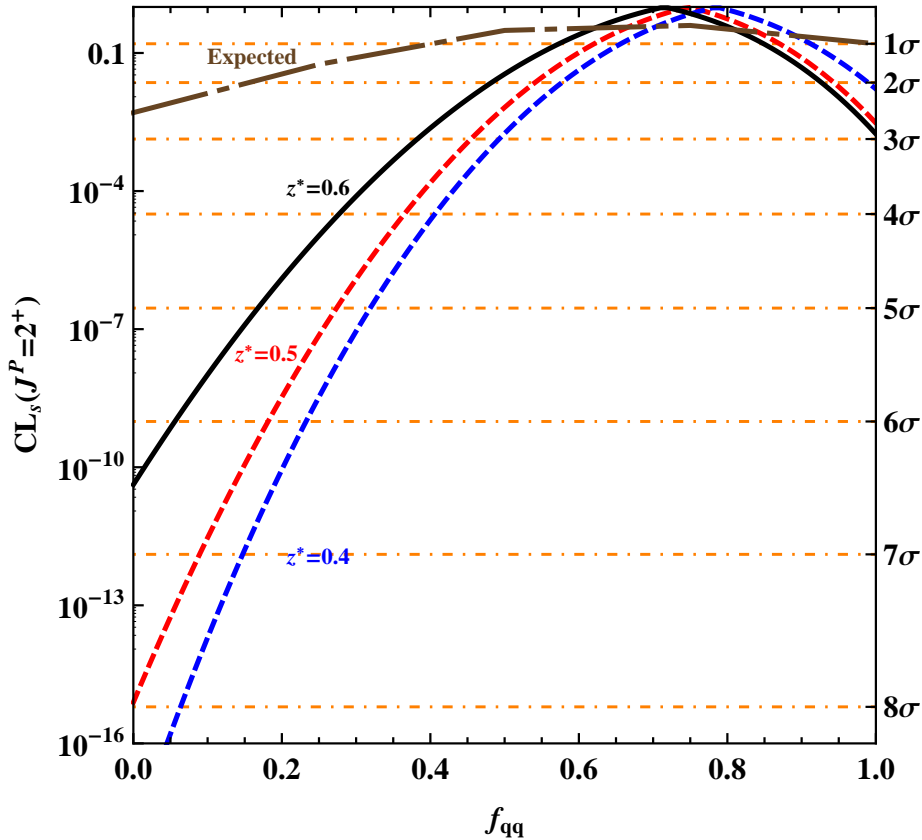}}
\caption{\label{fig4} The confidence level, ${\rm
CL_s}(J^P=2^+)$, of the $J^P=2^+$ hypothesis as a function of the
fraction $f_{qq}$ for spin-2 particle production, obtained from the
center-edge asymmetry measure at different $z^*=$ 0.4, 0.5 and 0.6
in the $H\to\gamma\gamma$ channel at the LHC with $\sqrt s=8$ TeV and
luminosity 20.7~${\rm fb}^{-1}$. On the right vertical axis, the
corresponding number of Gaussian standard deviations is given.
The dash-dotted line represents the expected confidence level from
the ATLAS analysis summarized in Table 4 of
Ref.~\cite{ATLAS:2013xla}.
}
\end{figure}

There is an alternative approach to quantify the separation power
by using the CLs prescription \cite{Read:2002hq}. The exclusion of the alternative spin-2
hypothesis in favor of the SM spin-0 hypothesis is evaluated in
terms of the corresponding ${\rm CLs}(J^P=2^+)$, defined as
\begin{equation}
{\rm CLs}(J^P=2^+) = \frac{p(J^P=2^+)}{1 - p(J^P=0^+)},
\label{CLS}
\end{equation}
where $p(J^P=2^+)$ is the $p$-value for spin-2 and
$p(J^P=0^+)$ is the $p$-value for spin-0, respectively. Spin-2
exclusion limits as functions of  $f_{qq}$ at three values of
$z^*=$0.4, 0.5 and 0.6 computed using the CLs prescription are shown
in Fig.~\ref{fig4}.\footnote{The numerical results
obtained for $f_{qq}=0$ and $z^*=0.5$ are consistent with those  presented in
Refs.~\cite{Alves:2012fb,Ellis:2012jv}. Also, one
should note that no shape systematic uncertainties other than PDF were
taken into account when evaluating the $A_{\rm CE}$ performance.}

\begin{figure}[tbh!] 
\vspace*{0.1cm} \centerline{ \hspace*{-2.0cm}
\includegraphics[width=12.0cm,angle=0]{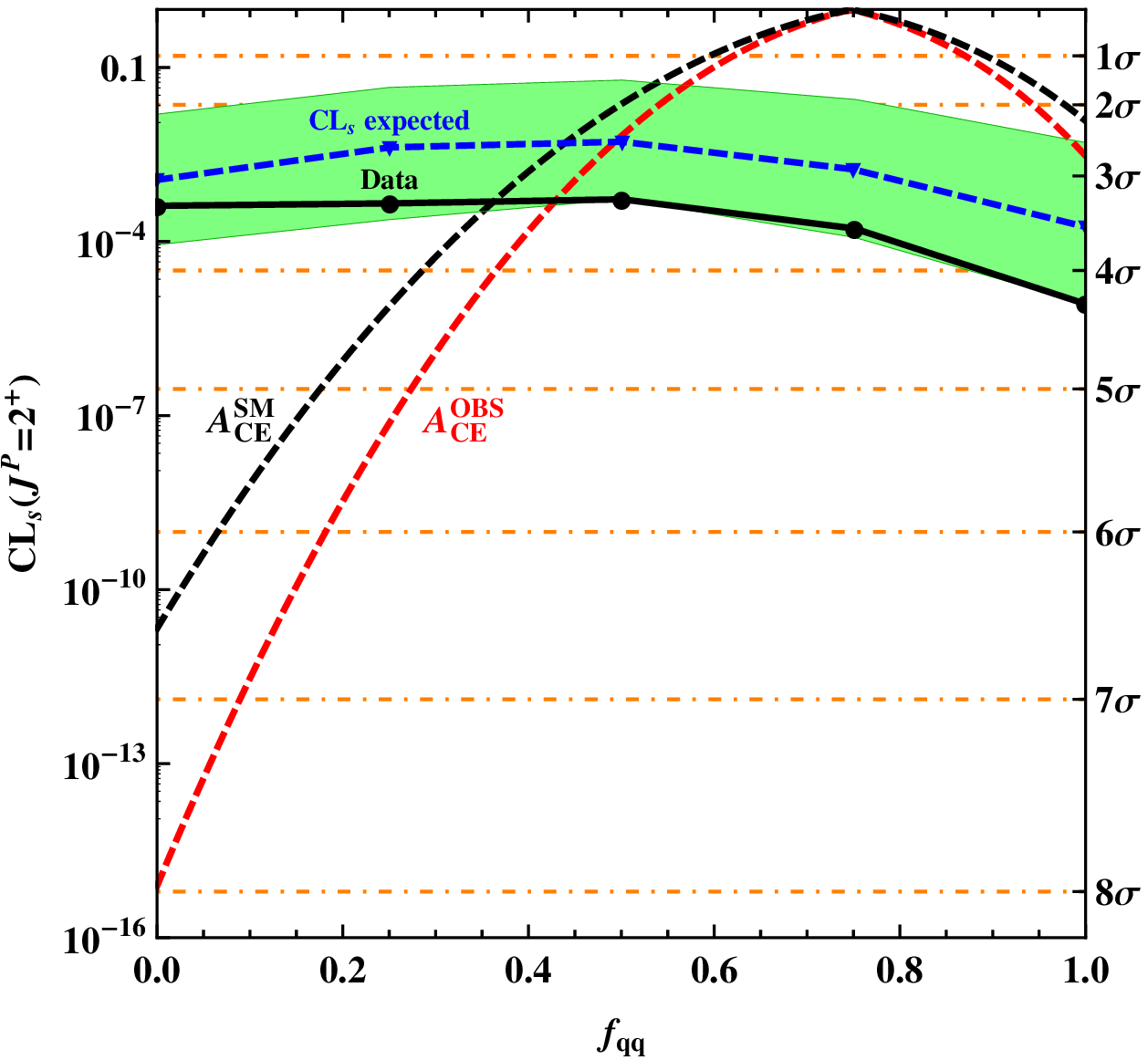}}
\caption{\label{fig5}
Comparison of the expected and observed
confidence levels, ${\rm CL_s}(J^P=2^+)$, of the $J^P=2^+$
hypothesis as functions of the fraction $f_{qq}$ for spin-2
particle production.
Expected (blue triangles/dashed line) and observed (black circles/solid
line) confidence levels are based on the experimental data and
obtained from the
combination of the angular distributions of the three channels
$H\to\gamma\gamma$, $H\to ZZ^*$ and  $H\to WW^*$ at $\sqrt s=7$ and 8
TeV  with ATLAS \cite{Aad:2013xqa}.
The green band represents the 68\% expected exclusion range
for a signal with assumed $J^P = 0^+$. On the right vertical axis, the
corresponding
number of Gaussian standard deviations is given.
The center-edge asymmetry measure, applied to the $H\to\gamma\gamma$ channel
at $z^*=0.5$, yields the expected ($A_{\rm CE}^{\rm SM}$) 
and observed ($A_{\rm CE}^{\rm OBS}$) confidence levels shown as dashed black and red curves. }
\end{figure}

It is instructive to compare the confidence level,
obtained in the present analysis with those available from the ATLAS
study of the three channels $H\to\gamma\gamma$, $H\to ZZ^*$ and $H\to
WW^*$ at $\sqrt s=8$ TeV and luminosity 20.7 ${\rm fb}^{-1}$
\cite{Aad:2013xqa}. Fig.~\ref{fig5} shows that $A_{CE}$
measurements are able to substantially increase the observed
confidence level, in particular in the range of parameter space
where $0<f_{qq} < 0.4$. In other words, in this range of
$f_{qq}$, $A_{CE}$ provides quite competitive information on the
spin of the Higgs-like boson with respect to that derived from the
commonly used analysis of angular distributions.

The result obtained from the $A_\text{CE}$ observable obviously depends on the number of events in this channel. Since the signal strength observed by ATLAS has recently been somewhat reduced, due in part to an improved photon energy calibration \cite{Aad:2014eha} and diphoton mass resolution, we show in Fig.~\ref{fig5} two curves: one corresponding to the observed angular distribution \cite{Aad:2013xqa,ATLAS:2013xla} (670 events, labeled $A_\text{CE}^\text{OBS}$) and one corresponding to the SM expectation (390 events,  labeled $A_\text{CE}^\text{SM}$).

\section{Concluding remarks} \label{sect:concl}
We have studied the possibility to determine the spin of the Higgs-like
boson with the center-edge asymmetry in the $H\to\gamma\gamma$ channel at ATLAS
with 8 TeV and integrated luminosity of $20.7~\text{fb}^{-1}$. In the present
analysis we compared the spin-0 hypothesis of the Higgs-like boson with that
of a graviton-like,  spin-2, particle with minimal couplings, taking into
account the possibility that the tensor particle might be produced via
quark-antiquark annihilation or gluon fusion.  We obtained the discrimination power as a function of two parameters,
the dynamical one $f_{qq}$ that determines the fraction of the $q\bar{q}$ mode
in the resonance production, and the kinematical one $z^*$ that defines
the center-edge asymmetry.

Optimization of the separation significance
on the kinematical parameter $z^*$ at
different $f_{qq}$ allows to find the region in the parameter plane where
the center-edge asymmetry could provide quite competitive information on the
spin of the Higgs-like boson with respect to that which is derived
from the more common angular-distribution analysis.
We found that
$A_\text{CE}$  provides discrimination between the scalar and tensor hypotheses
with ${\rm CL_s}<10^{-6}$ at $f_{qq}=0$ and
$z^*\approx 0.4$, a value that substantially exceeds the ATLAS expectations. For
increasing values of $f_{qq}$, the expected separation between spin-0 and spin-2
hypotheses is reduced, reaching a minimum at $f_{qq}\approx 0.75$ where
separation is impossible.
At higher energies, however, the gluon-gluon contribution would tend to increase, thus strengthening the usefulness of $A_\text{CE}$.

\section*{Acknowledgments}
A.A.\ Pankov and A.V.\ Tsytrinov would like to thank Nello Paver for valuable
discussions.
This research has been partially supported by the Abdus Salam ICTP
(TRIL and Associate Programmes),  the Collaborative Research
Center SFB676/1-2006 of the DFG at the Department of Physics of
the University of Hamburg and the Belarusian Republican Foundation
for Fundamental Research. The work of PO has been supported by the
Research Council of Norway.

\end{document}